\documentclass[twocolumn]{aastex6}
\usepackage{graphicx}	% Including figure files
\usepackage{amsmath}	% Advanced maths commands
\usepackage{amssymb}	% Extra maths symbols
\usepackage{booktabs}	% Pretty tables for all!
\usepackage{upgreek}
\usepackage{url}

\begin{document}

\title{A planet in an 840-d orbit around a \textit{Kepler} main-sequence A star found from phase modulation of its pulsations}

\author{Simon J. Murphy\altaffilmark{1} and Timothy R. Bedding}
\affil{Sydney Institute for Astronomy (SIfA), School of Physics, The University of Sydney, Australia\\
Stellar Astrophysics Centre, Department of Physics and Astronomy, Aarhus University, 8000 Aarhus C, Denmark}

\author{Hiromoto Shibahashi}
\affil{Department of Astronomy, The University of Tokyo, Tokyo 113-0033, Japan}

\altaffiltext{1}{simon.murphy@sydney.edu.au}

\begin{abstract}

We have detected a 12\,M$_{\rm Jup}$ planet orbiting in or near the habitable zone of a main-sequence A star via the pulsational phase shifts induced by orbital motion. The planet has an orbital period of $840\pm20$\,d and an eccentricity of 0.15. All known planets orbiting main-sequence A stars have been found via the transit method or by direct imaging. The absence of astrometric or radial-velocity detections of planets around these hosts makes ours the first discovery using the orbital motion. It is also the first A star known to host a planet within 1$\sigma$ of the habitable zone. We find evidence for planets in a large fraction of the parameter space where we are able to detect them. This supports the idea that A stars harbor high-mass planets in wide orbits.

\end{abstract}


%%%%%%%%%%%%%%%%%%%%
%%%%%%%%%%%%%%%%%%%%
%%%%%%%%%%%%%%%%%%%%

\section{Introduction}
\label{sec:intro}

The Kepler Space Telescope is now the most successful planet-finding mission to date. Most of its thousands of planets (and planet candidates) have been found around cool stars \citep{mullallyetal2015}, while exoplanets orbiting A stars remain elusive. After the announcement of an additional 1284 \textit{Kepler} exoplanets in 2016 May \citep{mortonetal2016}, still fewer than 20 A stars have confirmed planets.

Radial velocity observations of cool sub-giants, most notably via the `Retired A stars' project \citep{johnsonetal2007}, suggest that planet occurrence rates reach a maximum for stars of $1.9^{+0.1}_{-0.5}$\,M$_{\odot}$ \citep{reffertetal2015}. This coincides remarkably with the masses of main-sequence A stars in the classical instability strip, where delta Scuti ($\delta$\,Sct) pulsators are common. The apparent planet deficit around these main-sequence stars can be explained as an observational selection effect, caused by problems in the application of the most successful planet-hunting methods to these types of stars. The transit method has difficulty because of the pulsational luminosity variations, which amount to several mmag, and because planets occupy wider orbits around A stars (\citealt{johnsonetal2011}, cf. \citealt{lloyd2011}), resulting in a lower transit probability. The radial velocity method, on the other hand, is particularly hindered by the nature of A-type spectra. A stars are typically fast rotators, with the mean of the equatorial rotational velocity distribution exceeding 100\,km\,s$^{-1}$ \citep{abtandmorrell1995,royeretal2007}. Their high effective temperatures lead to fewer, shallower absorption lines, and these lines can be distorted by pulsation. Therefore the wavelengths of their spectral lines are not a precise standard of measure, and the state of the art is limited to 1--2\,km\,s$^{-1}$ precision \citep{beckeretal2015}.

Fortunately, the same pulsations that limit RV and transit surveys of A stars can themselves be used as precise clocks for the detection of orbital motion \citep{murphyetal2014}. The pulsations of $\delta$\,Sct stars are particularly well suited to this task \citep{comptonetal2016}.

\subsection{Known planets around A stars}

No planets have yet been discovered orbiting main-sequence A stars via the motion of their host stars, namely, via the RV, astrometric, or pulsation-timing methods.

Discoveries have been made using two other methods: transits and direct imaging. The former category includes \textit{Kepler} planet hosts (Kepler-13, 462, 516, 959, 1115, 1171 and 1517; \citealt{mortonetal2016}), plus WASP-33 \citep{colliercameronetal2010}, HAT-P-57 \citep{hartmanetal2015} and KELT-17 \citep{zhouetal2016}. The orbital periods of these systems are all under 30\,d, except for Kepler-462\,b whose period is 85\,d. These planets are strongly irradiated and would have high surface temperatures.

The imaged planets in the NASA Exoplanet Archive have very wide orbits. These include three objects (HD\,100546\,b, HIP\,78530\,b and $\kappa$\,And\,b) with masses above the canonical 13-M$_{\rm Jup}$ boundary between planets and low-mass brown dwarfs, which orbit B-type stars. Exoplanets orbiting A stars include Fomalhaut\,b \citep{stapelfeldtetal2004,chiangetal2009}, $\beta$\,Pic\,b \citep{lagrangeetal2010}, HD\,95086\,b \citep{rameauetal2013}, the multi-planet system HR\,8799\,b,c,d,e \citep{maroisetal2008}, and the planet in the triple system HD\,131399\,Ab \citep{wagneretal2016}, all of which have orbits ranging from 9 to 177 au.

Finally, there are some planets orbiting evolved hot stars, such as those discovered from eclipse timing variations of the white-dwarf binary NN\,Ser \citep{brinkworthetal2006,beuermannetal2010,parsonsetal2010}, the companion to the sdB star V0391\,Peg from pulsation timing variations \citep{silvottietal2007}, and companions to other sdB stars KIC\,10001893 \citep{silvottietal2014} and the now-doubtful case of KIC\,5807616 (\citealt{charpinetetal2011}, cf. \citealt{krzesinski2015}) from orbital brightness modulations.

In this paper we present the first planet discovered around a main-sequence A star by the motion of the host star, and the first to be in or near the habitable zone. We describe the observations and our method in Sect.\,\ref{sec:obs}. The planet's habitability and the implications for planet occurrence posed by this discovery are discussed in Sect.\:\ref{sec:habitability}.

\section{Observations and Analysis}
\label{sec:obs}

The host star, KIC\,7917485, was observed for the full four years of the \textit{Kepler} mission in long-cadence mode (30-min sampling). Our analysis used the multi-scale MAP data reduction of this data set \citep{smithetal2012,stumpeetal2012}. The time series spans 1461\,d with a duty cycle of 91\:per\:cent.

\begin{figure}
\begin{center}
\includegraphics[width=0.49\textwidth]{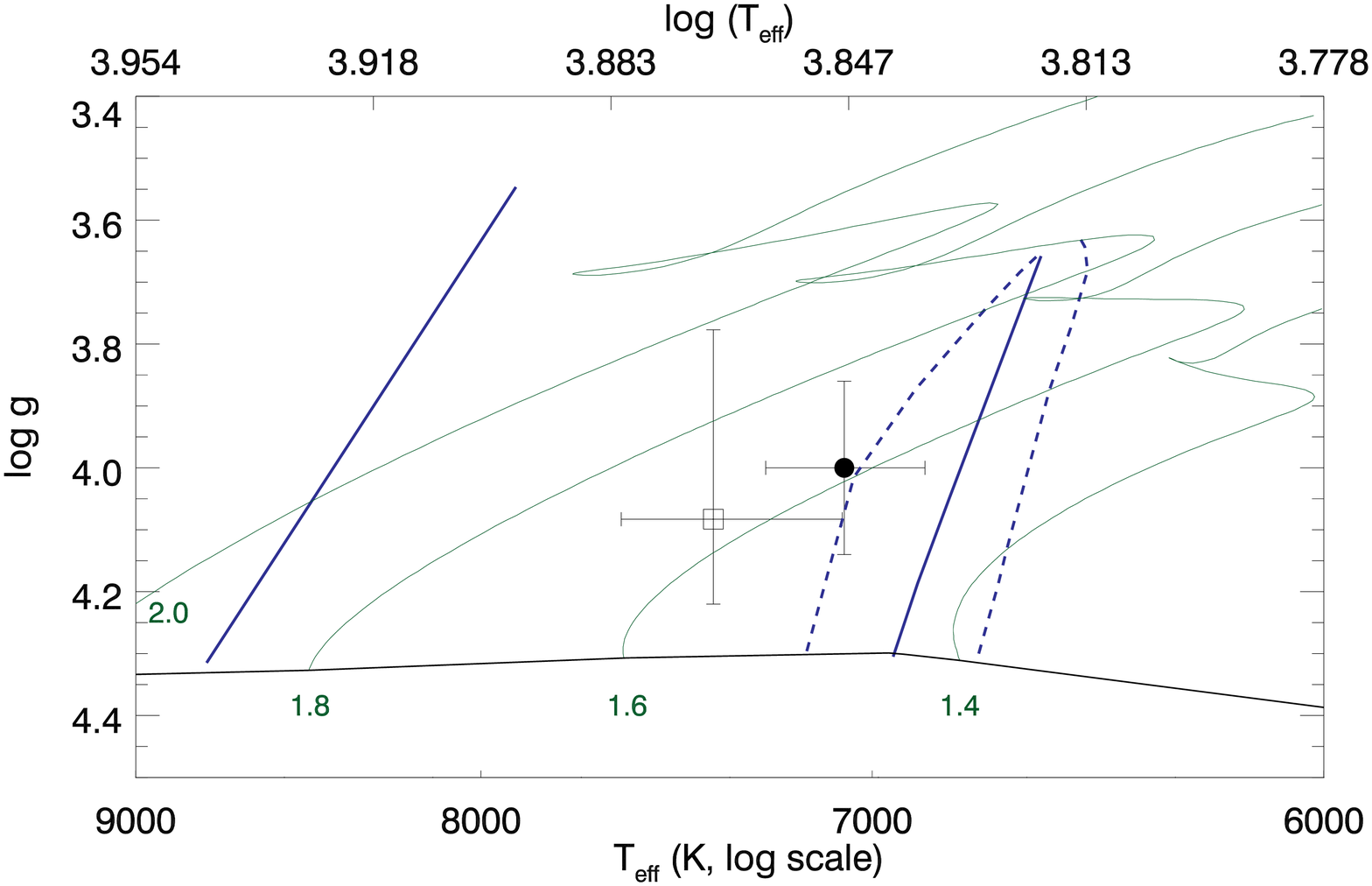}
\caption{Position of KIC\,7917485 on a $T_{\rm eff}$--$\log g$ diagram, according to LAMOST spectroscopy (filled circle) and revised KIC photometry (open square). The solid and dashed blue lines delineate the $\delta$\,Sct and $\gamma$\,Dor instability strips, respectively. Evolutionary tracks of solar metallicity and mixing length $\alpha_{\rm MLT} = 1.8$ are also shown \citep{grigahceneetal2005}, with corresponding masses in M$_{\odot}$ written where the tracks meet the zero-age main-sequence (black line). The peak of the planet occurrence rate distribution, at $1.9^{+0.1}_{-0.5}$\,M$_{\odot}$ \citep{reffertetal2015}, crosses the center of the $\delta$\,Sct instability strip.}
\label{fig:tlogg}
\end{center}
\end{figure}

This $V=13.2$\,mag star was observed spectroscopically with LAMOST \citep{decatetal2015}. The stellar atmospheric parameters were extracted by \citet{frascaetal2016}, showing KIC\,7917485 to be a main-sequence A star, in agreement with the photometric characterization by \citet{huberetal2014}. Its position on a $T_{\rm eff}$--$\log g$ diagram is shown in Fig.\,\ref{fig:tlogg}, and atmospheric parameters are given in Table\:\ref{tab:atmos}. The star is located in the $\delta$\,Sct instability strip and has a mass of approximately 1.63\,M$_{\odot}$.

\begin{table}
\caption{Global atmospheric parameters from LAMOST spectroscopy \citep{frascaetal2016} and revised KIC photometry \citep{huberetal2014}. We evaluated these quantities against evolutionary tracks to calculate a mass and luminosity, which are given below the mid-table break.}
\label{tab:atmos}
\begin{center}
\begin{tabular}{c c r@{ $\pm$ }l c}
\toprule
Parameter & Units & \multicolumn{2}{c}{Spectroscopy} & Photometry \\
\midrule
\vspace{1.5mm}$T_{\rm eff}$ & K & 7067 & 192 & $\phantom{-}7390^{+236}_{-318}$ \\
\vspace{1.5mm}$\log g$ & (cgs) & 4.00 & 0.14 & $\phantom{-}4.08^{+0.14}_{-0.31}$ \\
\vspace{1.5mm}$[$Fe/H$]$ & & $-0.02$ & $0.15$ & $-0.26^{+0.30}_{-0.36}$ \\
$v\sin i$ & km\,s$^{-1}$ & \multicolumn{2}{c}{$<$120} & \\
\midrule
\vspace{1.5mm}
$M_1$ & M$_{\odot}$ & \multicolumn{2}{c}{$1.63^{+0.15}_{-0.12}$} & $1.67^{+0.20}_{-0.12}$\\
$L_1$ & L$_{\odot}$ & \multicolumn{2}{c}{$9.9^{+4.8}_{-3.3}$} & $10.1^{+13.3}_{-\phantom{0}3.1}$ \\ 
\bottomrule
\end{tabular}
\end{center}
\end{table}

\begin{figure}
\begin{center}
\includegraphics[width=0.49\textwidth]{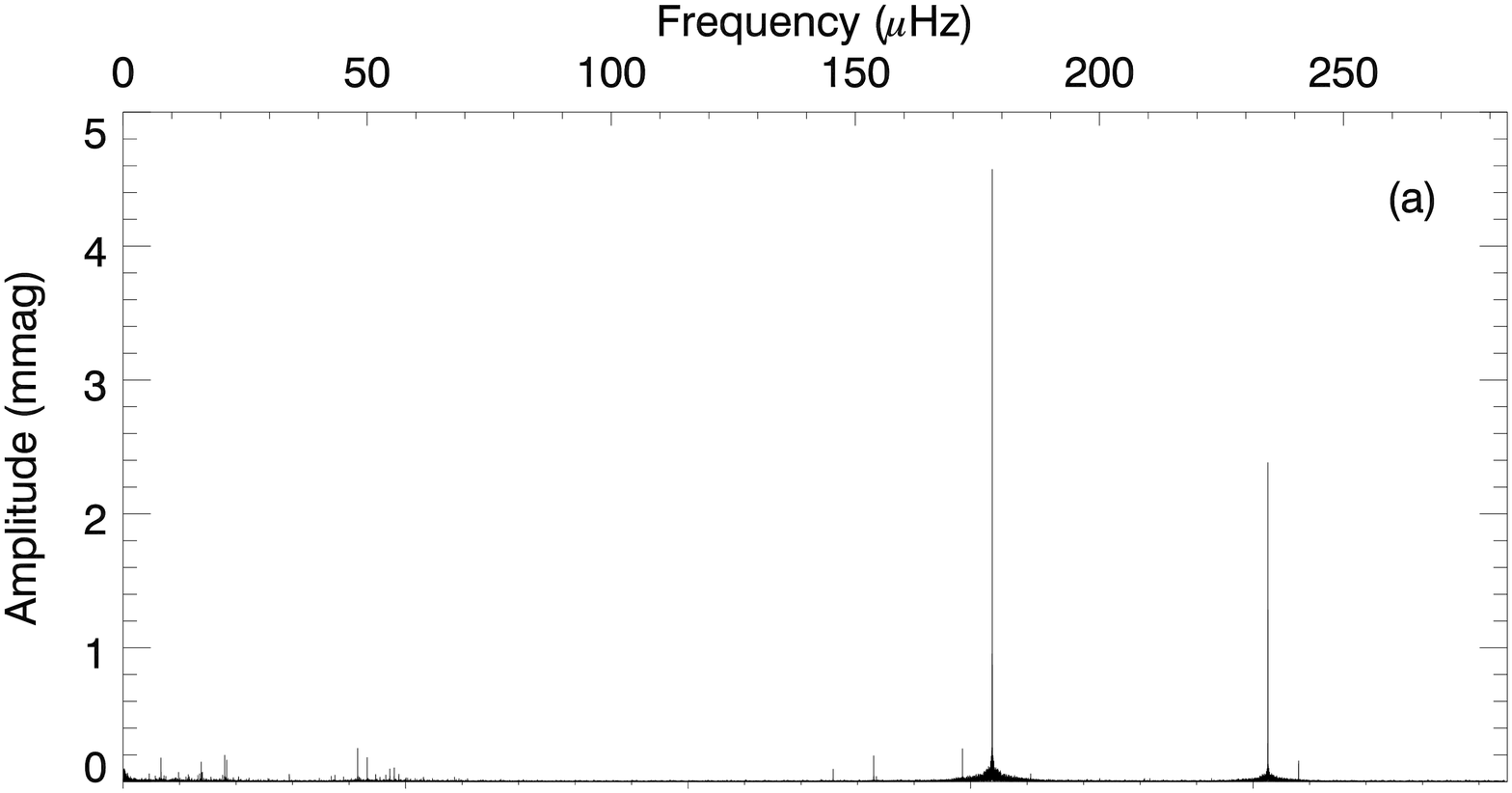}
\includegraphics[width=0.49\textwidth]{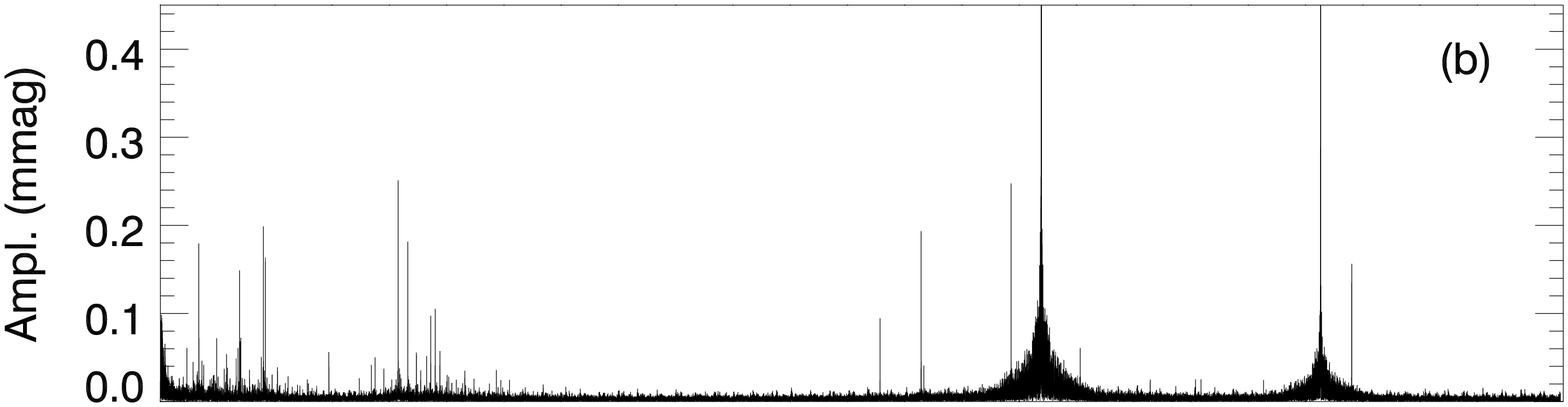}
\includegraphics[width=0.49\textwidth]{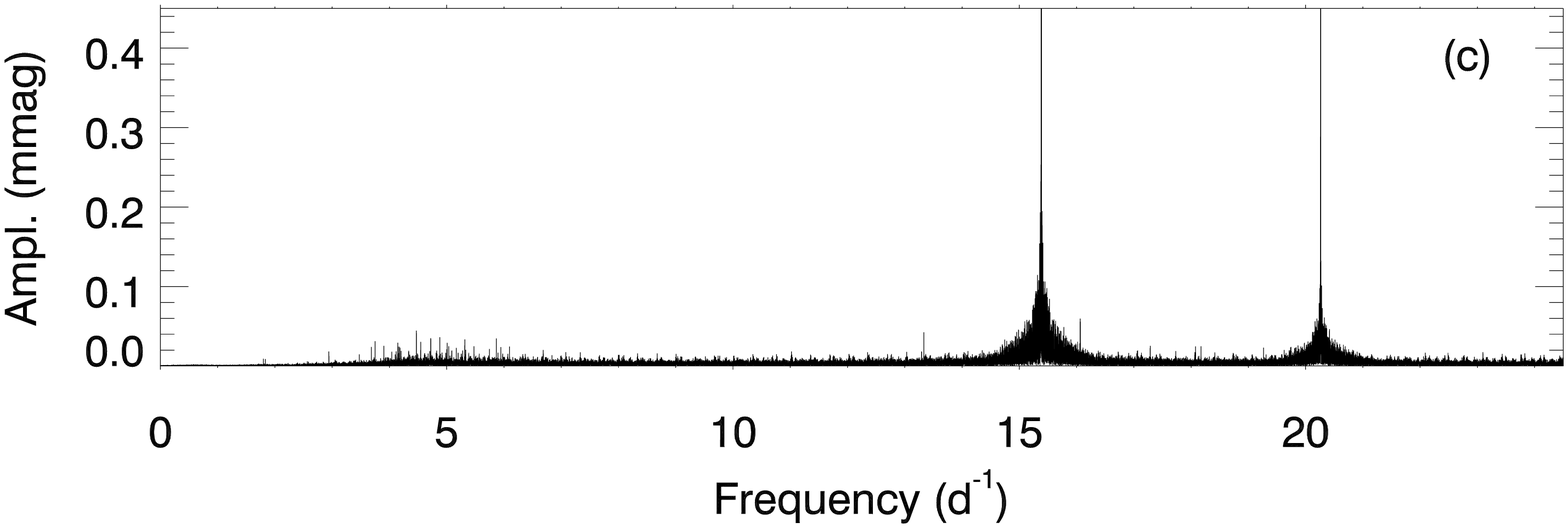}
\caption{The Fourier transform of the light curve of KIC\,7917485 before processing (a), also shown with vertical zoom (b). Panel (c) shows the high-pass-filtered data, with low-amplitude oscillations removed down to a 50-$\upmu$mag level. The horizontal axis is the same in each panel.}
\label{fig:ft}
\end{center}
\end{figure}

We detected the planet by phase modulation (PM) of the stellar pulsations \citep{murphyetal2014}. A Fourier transform of the light curve (Fig.\,\ref{fig:ft}a) is dominated by two oscillation modes at frequencies $f_1 = 15.3830026(1)$ and $f_2 = 20.2628968(3)$\,d$^{-1}$, where the uncertainty on the final digit has been given in parentheses. These two modes were used in the PM analysis. Other significant peaks have amplitudes that are at least an order of magnitude lower. Their signal-to-noise ratios are too low to add usefully to the PM analysis, but their presence adds unwanted variance to the data. We therefore subtracted all peaks above 50\,$\upmu$mag from the data, except for $f_1$ and $f_2$, by fitting their frequencies to the light curve with a nonlinear least-squares algorithm. We also high-pass-filtered the light curve to remove any remaining instrumental signal and low-frequency oscillations, preserving all content at frequencies above 5\,d$^{-1}$. The Fourier transform after the additional processing is shown in Fig.\,\ref{fig:ft}c.

\begin{figure}
\begin{center}
\includegraphics[width=0.49\textwidth]{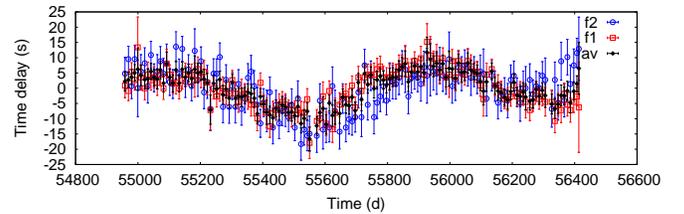}
\caption{Delays in the light arrival time for the independent oscillation frequencies $f_1$ (red squares) and $f_2$ (blue circles). Error bars are the formal least-squares uncertainties. Weighted averages of the two measurements in each 10-d segment are indicated as black diamonds.}
\label{fig:tds}
\end{center}
\end{figure}

We divided the light curve into 10-d segments to look for shifts in the phases of $f_1$ and $f_2$. These phases were converted into delays in the light arrival time (`time delays') following the method of \citet{murphyetal2014}. Time delays of $f_1$ and $f_2$ show identical periodic variation (Fig.\,\ref{fig:tds}), which we attribute to a sub-stellar companion. Values for the orbital parameters were initially obtained using formulae from \citet{murphyandshibahashi2015}, and then refined with an MCMC algorithm. The MCMC analysis used a Metropolis--Hastings algorithm \citep{metropolisetal1953,hastings1970} with symmetric proposal distributions based on gaussian-distributed random numbers. Trial runs were made to determine appropriate step sizes in each of the five orbital parameters fitted: the orbital period, $P_{\rm orb}$, projected light travel time across the orbit, $a_1 \sin i / c$, eccentricity, $e$, the phase of periastron passage calculated relative to the first time delay observation, $\phi_{\rm p}$, and the angle of the ascending node, $\varpi$.\footnote{We use $\varpi$, rather than $\omega$, because of the common use of $\omega$ to represent angular oscillation frequencies in asteroseismology.} We fitted a linear term to the time delays as an additional parameter to correct for slight inaccuracies in the oscillation frequencies. Proposed steps in all six parameters were made simultaneously. A total of 200\,000 steps were made, with a final acceptance rate of 0.20. For further details on the methodology, including its verification by radial velocities, see \citet{murphyetal2016}.

\begin{figure}
\begin{center}
\includegraphics[width=0.49\textwidth]{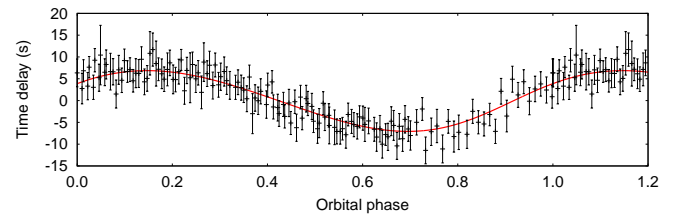}
\caption{Comparison of the weighted average time delays with the best-fitting orbital parameters from the MCMC analysis (given in Table\:\ref{tab:orbit}).}
\label{fig:tdmcmc}
\end{center}
\end{figure}

Table~\ref{tab:orbit} gives the best-fitting orbital parameters, which are compared with the observations in Fig.\,\ref{fig:tdmcmc}.

\begin{table}
\centering
\caption{Orbital parameters for the KIC\,7917485 system. The value of $\phi_{\rm p}$ is calculated with respect to the first time-delay measurement at $t_0 ({\rm BJD}) = 2\,454\,958.39166.$
$M_1$ is inferred non-dynamically, from the spectroscopic parameters, which allows $M_2 \sin i$ and $a_2 \sin i$ to be calculated. $K_1 \sin i$, the projected radial velocity semi-amplitude, is calculated from the orbital parameters and provided for reference; it is not used to derive the orbital solution.}
\label{tab:orbit}
\begin{tabular}{c c c}
\toprule
\multicolumn{1}{c}{Parameter} & Units & \multicolumn{1}{c}{Values}\\
\midrule
\vspace{1.0mm}
$P_{\rm orb}$ & d & $840^{+22}_{-20} $\\
\vspace{1.0mm}
$e$ & & $0.15^{+0.13}_{-0.10}$ \\
\vspace{1.0mm}
$\phi_{\rm p}$ & $[$0--1$]$ & $0.89^{+0.10}_{-0.12}$\\
\vspace{1.0mm}
$f(m_1,m_2,\sin i)$ & M$_{\odot}$ & $5.3^{+1.1}_{-1.0}\times10^{-7}$\\
\midrule
\multicolumn{3}{c}{\it Star}\\
\midrule
\vspace{1.0mm}
$M_1$ & M$_{\odot}$ & $1.63^{+0.15}_{-0.12}$ \\ 
\vspace{1.0mm}
$a_1 \sin i / c$ & s & $7.1^{+0.5}_{-0.4}$\\
\vspace{1.0mm}
$\varpi$ & rad & $3.0^{+0.6}_{-0.7}$ \\
\vspace{1.0mm}
$K_1 \sin i$ & m\,s$^{-1}$ & $186^{+17}_{-13}$ \\
\midrule
\multicolumn{3}{c}{\it Planet}\\
\midrule
\vspace{1.0mm}
$M_2 \sin i$ & M$_{\odot}$ & $0.0113^{+0.0008}_{-0.0006}$ \\ 
\vspace{1.0mm}
  & M$_{\rm Jup}$ & $11.8^{+0.8}_{-0.6}$ \\ 
\vspace{1.0mm}
$a_2 \sin i / c$ & s & $1017^{+136}_{-110}$\\
\vspace{1.0mm}
 $a_2 \sin i$ & au & $2.03^{+0.27}_{-0.22}$\\
\vspace{1.0mm}
$\varpi$ & rad & $6.1^{+0.6}_{-0.7}$ \\
\bottomrule
\end{tabular}
\end{table}

\section{Discussion}
\label{sec:habitability}

\subsection{Habitable zone}

The minimum mass of the sub-stellar companion to KIC\,7917485 (11.8\,M$_{\rm Jup}$) is close to the planet--brown-dwarf mass boundary. Although such a body would not be expected to have a solid surface, it could presumably host smaller exomoons. It is therefore interesting to determine whether the planet lies within the habitable zone.

The habitable zone is broader for A stars than for cooler stars because the inner edge extends to higher incident surface fluxes for hotter stars \citep{kopparupuetal2014}. Their calculations indicate that, at the temperature of KIC\,7917485, the inner edge lies between $\sim$1.2 and 1.4 times the surface flux at Earth, depending on the mass of the potentially habitable body.

The mean projected separation between the components, $a \sin i = a_1 \sin i + a_2 \sin i$, is 2.05\,au. At this separation, the luminous A star irradiates the companion to a surface flux ratio, $S/S_0$, of 2.36 times the flux at Earth. Statistical correction for random inclination gives $S/S_0 = 1.76$ as the most probable value.

The luminosity of the A star is not well constrained, and is a strong function of main-sequence age. At the $1\sigma$ lower luminosity limit (see Table\:\ref{tab:atmos}), those values of $S/S_0$ reduce to 1.72 and 1.27, respectively. Thus the position of KIC\,7917485b is consistent with the habitable zone at the $1\sigma$ level. The companion would have been closer to the center of the habitable zone earlier in the star's lifetime. The luminosity will be refined substantially when the {\it Gaia} mission provides a distance measurement, and the question of habitability can be reassessed.

\subsection{Implications for planet occurrence rates}
\label{sec:occurrence}

KIC\,7917485b is the least massive companion that we have found in \textit{Kepler} data with the PM method (\mbox{Murphy} et al.\ 2016, in preparation). We have also found two other stars with time delay variations consistent with planetary companions having periods longer than the 4-yr data set (KIC\,9700322, KIC\,8453431), but the finite duration of \textit{Kepler} time-series does not allow the orbits to be fully parametrized. These detections allow us to comment on the planet occurrence around A stars.

The detectability of low-mass companions is very sensitive to the noise in the Fourier transform of the time delays, which is determined by the pulsation properties \citep{murphyetal2016}. We quantified this noise level for 2040 pulsating single A stars and found that only five of them had lower noise levels than KIC\,7917485. Against the same sample, KIC\,9700322 and KIC\,8453431 ranked as the 9$^{\rm th}$ lowest and 2$^{\rm nd}$ lowest, respectively. In other words, we have been able to detect a planetary-mass companion in one of the nine stars with the lowest noise levels, and two others show variations that are consistent with planetary-mass companions. This fact is in strong support of existing observations that intermediate-mass stars (`retired A stars') tend to host high-mass planets in wide orbits \citep{johnsonetal2011}.

%%%%%%%%%%%%%%%%%%%%
%%%%%%%%%%%%%%%%%%%%
%%%%%%%%%%%%%%%%%%%%

\section{Conclusions}
\label{sec:conclusions}

We have found a planetary companion of \mbox{$M\sin i =11.8$\,M$_{\rm Jup}$} near the inner edge of the habitable zone of KIC\,7917485. This is the first planet orbiting a main-sequence A star to be discovered via the motion of its host star. Other planets orbiting A stars have been discovered via the transit method and have short periods, or been discovered by direct imaging and have very long periods. Our finding is particularly significant because no other method is presently capable of detecting non-transiting planets around these stars with periods of a few years, i.e., near their habitable zones.

KIC\,7917485 has particularly low noise levels in its light arrival time delays. We analyzed other stars with similarly low noise and for two of them we also found evidence for planetary-mass companions with periods similar to the 4-yr time span of \textit{Kepler} data. This fact strongly supports the idea that intermediate-mass stars tend to host high-mass planets in wide orbits.

\end{document}